\documentclass[aps,showpacs,floats]{revtex4}
\usepackage{amssymb}
\usepackage{graphicx}
\usepackage{bm}

\begin{document}

\bibliographystyle{apsrev}

\def\v#1{{\bf #1}}
\newcommand{\vq}{{\bf q}}
\newcommand{\vQ}{{\bf Q}}
\newcommand{\hq}{{\hat{q}}}
\newcommand{\vk}{{\bf k}}
\newcommand{\vR}{{\bf R}}
\newcommand{\lm}{\lambda}
\newcommand{\da}{\dagger}
\newcommand{\tu}{\tilde{u}}
\newcommand{\tv}{\tilde{v}}

\newcommand{\la}{\langle}
\newcommand{\ra}{\rangle}
\newcommand{\lc}{\lowercase}
\newcommand{\hg}{\hat{g}_0}

\newcommand{\no}{\nonumber}
\newcommand{\be}{\begin{equation}}
\newcommand{\ee}{\end{equation}}
\newcommand{\bea}{\begin{eqnarray}}
\newcommand{\eea}{\end{eqnarray}}

\newcommand{\SK}{RM$_4$X$_{12}$}
\newcommand{\LO}{L\lc{a}O\lc{s}$_4$S\lc{b}$_{12}$}
\newcommand{\PGe}{P\lc{r}P\lc{t}$_4$G\lc{e}$_{12}$}
\newcommand{\PO}{P\lc{r}O\lc{s}$_4$S\lc{b}$_{12}$}
\newcommand{\LR}{L\lc{a}R\lc{u}$_4$S\lc{b}$_{12}$}
\newcommand{\PR}{P\lc{r}R\lc{u}$_4$S\lc{b}$_{12}$}

\newcommand{\LOx}{L\lc{a}(O\lc{s}$_{1-x}$R\lc{u}$_x$)$_4$S\lc{b}$_{12}$}
\newcommand{\POx}{P\lc{r}(O\lc{s}$_{1-x}$R\lc{u}$_x$)$_4$S\lc{b}$_{12}$}
\newcommand{\POy}{L\lc{a}$_{1-y}$P\lc{r}$_y$O\lc{s}$_4$S\lc{b}$_{12}$}
\newcommand{\POxy}{L\lc{a}$_{1-y}$P\lc{r}$_y$(O\lc{s}$_{1-x}$R\lc{u}$_x$)$_4$S\lc{b}$_{12}$}

\title{Nonadiabatic effects of rattling phonons and 4f excitations in  \POx}

\author{ Peter Thalmeier}
\affiliation{Max Planck Institute for Chemical Physics of Solids, 01187 Dresden, Germany}

\begin{abstract}
In the skutterudite compounds the anharmonic 'rattling' oscillations of 4f-host ions in the  surrounding Sb$_{12}$ cages are found to have significant influence on the low temperature properties. Recently specific heat analysis of \POx~
has shown that the energy of crystalline electric field (CEF) singlet-triplet excitations increases strongly with Ru-concentration x and crosses the almost constant rattling mode frequency $\omega_0$ at about $x\simeq 0.65$. Due to magnetoelastic interactions this may entail prominent nonadiabatic effects in inelastic neutron scattering (INS) intensity and quadrupolar susceptibility. Furthermore the Ru- concentration dependence of the superconducting T$_c$, notably the minimum at intermediate x is explained as a crossover effect from pairforming aspherical Coulomb scattering to pairbreaking exchange scattering.
\end{abstract}

\pacs{63.20.kd, 75.10.Dg, 74.70.Tx, 74.20.-z}

\maketitle

\section{Introduction}
\label{sec:intro}

The tetrahedral rare earth skutterudite compounds \SK~ (R= rare earth, M = Ru, Fe, Os and  X = P, As , Sb) are found to exhibit a great variety of electronic ground  states driven by 4f electron correlations \cite{Aoki05}. Metallic heavy electron and mixed valent as well as Kondo semiconducting behaviour may be found. The 4f ground state may show quadrupolar or more exotic multipolar order \cite{Kuramoto09,Takimoto06,Shiina08}. In \PO~ and substituted compounds superconductivity appears \cite{Bauer02,Frederick04,Yogi06} which is possibly of strongly anisotropic nature \cite{Izawa03,Yogi06,Parker08}. The latter has recently also been found in \PGe \cite{Maisuradze09}.  It has been proposed \cite{Chang07,Koga06}  that low-energy crystalline electric field (CEF) excitations  play an important role in the formation of Cooper pairs and likewise in the quasiparticle mass enhancement \cite{Goremychkin04,Zwicknagl09}. These only slightly dispersive excitations were identified in INS experiments \cite{Kuwahara05} in \PO . Furthermore NMR/NQR \cite{Nakai08} and ultrasonic \cite{Goto05,Hattori07} measurements give evidence for the importance of  anharmonic 'rattling' oscillations of 4f host ions in the cages formed by surrounding Sb$_{12}$ icosahedrons on the low temperature properties.\\

Recent systematic specific heat measurements in the substitution series \POx~ have shown \cite{Miyazaki09}  that both CEF excitations and local rattling mode give contributions in addition to the usual Debye part. From the analysis Miyazaki et al were able to obtain the dependence of the low energy singlet-triplet CEF excitation $\Delta(x)$ and the rattling phonon frequency of the Pr host $\omega_0(x)$ on the Ru - concentration x. It was found  that $\Delta(x)$ increases with Ru content (x) from $\Delta(0) \simeq 8$ K  to $\Delta(1) \simeq 84$ K. For $x_c\simeq 0.65$ the triplet excitation energy crosses the oscillator energy $\omega_0\simeq 45$ K of the low energy anharmonic rattling phonon associated with the Pr oscillations in the cage. The latter is almost independent of the Ru concentration x. 
Earlier results \cite{Frederick04} showed that the critical temperature T$_c$(x) varies between T$_c$(0) = 1.85 K and  T$_c$(1) = 1.20 K and exhibits a minimum in between with T$_c$(x$_c$) = 0.7 K around the same Ru content x$_c$. This raises the question whether there is a connection between the crossing effect and the observed T$_c$ minimum and whether the nonadiabatic effects play a role in the appearance of the minimum. \\

In this work we discuss a model for the nonadiabatic effects between rattling phonons and CEF singlet-triplet excitations
in \POx. These effects should be primarily observable close to x$_c$ where an anticrossing and formation of 'vibronic'  or mixed modes is expected. The latter have partly a phononic and partly a CEF excitation nature. Although similar effects are known for other rare earth compounds \cite{Thalmeier82} they have sofar not been identified in the skutterudite family. The observed $\Delta(x)$, $\omega_0(x)$ behaviour in \POx~ however suggests their presence.
The mixed mode formation should lead to a direct clear signature in the dipolar INS cross section, magnetic and quadrupolar susceptibility as well as rattling phonon spectral function. Indirectly other physical quantities like NMR relaxation, resistivity and T$_c$ suppression or enhancement should also be influenced. Some of these effects will be discussed in the present work within an exactly solvable bosonic model for the two types of coupled excitations.\\

In Sec.~\ref{sec:model} the vibronic model is introduced and solved in Sec.~\ref{sec:vibron}. The  dynamical susceptibilities and associated structure functions, in particular the dipolar one relevant for INS are derived in Sec.~\ref{sec:dynsus}.
Furthermore we investigate the influence of vibronic excitations on the formation or breaking of Cooper pairs and the resulting T$_c$(x) variation with Ru content in Sec.~\ref{sec:crittemp}. Finally Sec.~\ref{sec:conc} gives the summary and conclusion.

\section{Model definition}
\label{sec:model}

First we consider the local 4f electronic part. The effect of a tetrahedral CEF on the Pr$^{3+}$ has been studied in detail in Ref.~\onlinecite{Takegahara01}. Its main consequence is a mixing of the cubic (O$_h$) $\Gamma_4$ and $\Gamma_5$ triplets to tetrahedral $\Gamma_4^{(1,2)}$ triplets which have both dipolar and quadrupolar transitions from the ground state singlet $\Gamma_1$. In the following we restrict to $\Gamma_1$ and lowest $\Gamma_4^{(2)}\equiv\Gamma_t$ triplet states as shown by Shiina \cite{Shiina04a}. 
The singlet-triplet CEF Hamiltonian in \POx~  may then be written in bosonic form as
\bea
H_{st}=\Delta(x)\sum_n(a_n^\dagger a_n+\frac{1}{2})\qquad(n=x,y,z)
\eea
where $\Delta(x)$ is the singlet-triplet CEF splitting of \POx~ that varies from $\Delta(0)=8$ K for the Os compound  to 
$\Delta(1)=84$ K for the Ru system. The triplet states are linear combinations of cubic  $\Gamma_5$ and $\Gamma_4$ states \cite{Takegahara01} described by
\bea
|\Gamma_tn\ra=(1-d^2)^\frac{1}{2}|\Gamma_5n\ra+d|\Gamma_4n\ra
\eea
where d characterizes the strength of the tetrahedral CEF part (Appendix \ref{sec:A1}). When the latter is small ($d\ll 1$) then  $\Gamma_t$ is close to the nonmagnetic cubic  $\Gamma_5$ triplet. When the tetrahedral CEF part dominates we have $d^2\rightarrow \frac{1}{2}$ (Eq.(\ref{eq:CEFmix})) and $\Gamma_t$ is an equal-amplitude mixture of nonmagnetic  $\Gamma_5$ and magnetic $\Gamma_4$ cubic triplets.
The $|\Gamma_tn\ra$ states are created by the bosonic operators a$_n$ (n=1-3)  according to $|\Gamma_tn\ra = a_n^\dagger|\Gamma_s\ra$ from the singlet ground state \cite{Shiina04a,Shiina04b}. They are related to $a_n$ ($n=x,y,z$) through $a_1=-(1/\sqrt{2})(a_x-ia_y);\; a_2=a_z;\; a_3=(1/\sqrt{2})(a_x+ia_y)$. The singlet-triplet system has quadrupolar and , because of the tetrahedral mixing amplitude d, also dipolar matrix elements for inelastic transitions. In bosonic representation the dipole operators are given by
\be
{\bf J}=b_D(a_x+a^\da_x, a_y+a^\da_y, a_z+a^\da_z)
\label{eq:dipolar}
\ee
where  $b_D=2\sqrt{\frac{5}{3}}d$ is the dipolar matrix element $\sim d$.
The $\Gamma_5$-type quadrupolar operators O$_n$ ($n=yz,zx,xy$) are generally given in terms of the total angular momentum components $J_n$ ($n = x,y,z$). In bosonic representation one has
\bea
O_{yz}&=&J_yJ_z+J_zJ_y=ib_Q(a_x-a_x^\dagger) \no\\
O_{xz}&=&J_xJ_z+J_zJ_x=ib_Q(a_y-a_y^\dagger)  \\
O_{xy}&=&J_xJ_y+J_yJ_x=ib_Q(a_z-a_z^\dagger) \no
\label{eq:quadru}
\eea
where $b_Q=\frac{2}{\sqrt{3}}\sqrt{35(1-d^2)}$ is the quadrupolar singlet-triplet matrix element which is maximal for d=0, contrary to $b_D$. These are the order parameters in the field-induced antiferroquadrupolar phase of \PO~\cite{Aoki02,Vollmer03,Tayama03,Rotundu04,Shiina04a} and their dynamics corresponds to the excitation of quadrupolar excitons \cite{Kuwahara05,Shiina04b}.

\begin{figure}
\includegraphics[width=8cm]{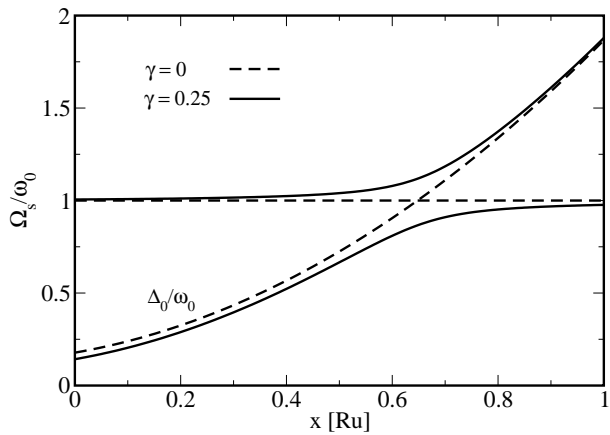}
\caption{Mode frequencies as function of Ru content x. At zero coupling $\gamma$ (dashed lines) flat rattling mode and increasing singlet-triplet CEF excitation cross around $x_c\simeq 0.65$. A finite coupling $\gamma$ leads to anti-crossing vibronic modes $\Omega_\phi$ (upper mode) and $\Omega_\psi$ (lower mode) around $x_c$. }
\label{fig:Fig1}
\end{figure}

Now we introduce the rattling phonon part which may be seen as a low frequency optical phonon corresponding to the anharmonic movement of the heavy Pr ion in the wide cage formed by the Sb$_{12}$ icosahedron. Such almost dispersionless rare earth host modes lying within the acoustic phonon bands are reported in Ref.~\cite{Lee06} for the Ce skutterudite. They belong to T$_1$ representation of $T_h$ and therefore are triply degenerate, corresponding to the three Cartesian directions of rattling motion in the cage.
Due to the anharmonic potential of the cage the effective rattling frequency $\omega_e(T)$ may be temperature dependent, similar as in the $\beta$- pyrochlore superconductor 
KOs$_2$O$_6$ \cite{Dahm07,Chang09}. On the other hand the low temperature effective rattling frequency $\omega_0=\omega_e(T=0)$ is almost independent of Ru content x with $\omega_0(x) \simeq 45$ K. Therefore, as observed in Ref.~\onlinecite{Miyazaki09} the singlet- triplet energy $\Delta(x)$ crosses the rattling frequency around $x_c\simeq 0.65$ (dashed lines in Fig.~\ref{fig:Fig1}) which is, incidentally, close to the Ru concentration where T$_c(x)$ shows its minimum. In the quasiharmonic approximation \cite{Dahm07} the rattling phonon part at low temperature is given by
\bea
H_r=\omega_0\sum_n(b_n^\dagger b_n+\frac{1}{2})
\eea
Where n=1-3 denotes one of the triply degenerate host modes which are created by the phonon operators $b_n^\dagger$.

The coupling of rattling modes and local CEF excitations (all dispersive effects in phonons and CEF excitations are neglected) may be written in terms of displacements and 4f quadrupoles as \cite{Thalmeierbook} 
\bea
H_{r-4f}=g_0Q_0\sum_{in}(b_n(i)+b_n^\dagger(i))O_n(i)
\eea
where $Q_0=(2MN\omega_0)^{-\frac{1}{2}}$ (M=mass of Pr and N=number of sites i) and g$_0$ is the coupling constant. Expressing the quadrupole operators with singlet-triplet boson operators the total Hamiltonian for each site is given by 
\bea
H=\sum_n[\frac{1}{2}\omega_0(b_n^\da b_n + b_n b_n^\da )+ \frac{1}{2}\Delta( a_n^\da a_n + a_n a_n^\da)
+i\tilde{g}_0b_Q(b_n a_n - b_n^\da a_n^\da + b_n^\da a_n - b_n a_n^\da)] 
\eea
This is the bosonic model Hamiltonian used in the following analysis. The first part describes three degenerate rattling phonon modes the second singlet-triplet CEF exitations and the last one their local magnetoelastic interactions.

It is bilinear in the singlet-triplet CEF (a) and phononic (b) boson operators and may therefore be diagonalised. For that purpose we write it in matrix form as 
\be
H=\sum_n
\left(\matrix{
a_n^\da & b_n^\da & a_n & b_n}\right)
\left(\matrix
{\frac{\Delta}{2}  & -i\hg & 0 &-i\hg \cr
i\hg & \frac{\omega_0}{2} & -i\hg & 0\cr
0& i\hg & \frac{\Delta}{2} & i\hg \cr
i\hg& 0 & -i\hg &\frac{\omega_0}{2}}\right)
\left(\matrix{
a_n\cr
b_n\cr
a^\da_n\cr
b^\da_n}\right)
\label{eq:ham1}
\ee

Here we defined $\tilde{g}_0=g_0(2M\omega_0)^{-\frac{1}{2}}$ and $\hg = b_Q\tilde{g}_0$.

\section{Vibronic excitations}
\label{sec:vibron}
%
%
The eigenstates of the model Hamiltonian in Eq.~(\ref{eq:ham1}) are the local vibronic modes of Pr$^{3+}$, i.e., mixed rattling phonon and singlet-triplet CEF excitations. The mode mixing becomes strong close to the crossing of $\Delta(x)$ with $\omega_0$ at $x_c$. The formation of vibronic modes therefore should influence physical properties, in particular close $x_c$.
To calculate the vibronic modes we express the model Hamiltonian in $2\times2$ block form: 
\be
H=\sum_{n =x,y,z}
\left(\matrix{
\alpha_n^\da & \alpha_n}\right)
\left(\matrix
{D_{1n} & D_{2n} \cr
D_{3n} & D_{4n}}\right)
\left(\matrix{
\alpha_n\cr
\alpha^\da_n}\right)
\label{eq:ham2}
\ee
where $\alpha^\da_n=(a^\da_n b^\da_n)$ and  $D_{3n}=D^\da_{2n}$. The D-matrices are the $2\times 2$ blocks in Eq.~(\ref{eq:ham1}). This quadratic form can be diagonalized by a generalised Bogoliubov or paraunitary transformation $\theta$ \cite{Colpa78} with the property $\theta I\theta^\da = I$ where $I = diag(11-1-1)$. In the following the mode degeneracy index ($n =x,y,z$) will be suppressed.
Applying the paraunitary transformation we get $D\theta^{-1}=I\theta^{-1}L$ where the column vectors $w_\rho (\rho=1-4)$ of $\theta^{-1}$ are  the eigenvectors of the equation $(ID-\lm_\rho1)w_\rho=0$ corresponding to the eigenvalue $\lm_\rho$ determined by the secular equation $det(D-\lm_\rho I)=0$. Furthermore $L= diag(\lm_+,\lm_-,-\lm_+,-\lm_-)$. The vibronic eigenvalues can be obtained as
\bea
\label{eq:eigen1}
\lm_\pm&=&[\epsilon_0\pm R]^\frac{1}{2}\no\\
\epsilon_0&=&\frac{1}{2}\Bigl(\frac{\omega_0^2}{4}+\frac{\Delta_0^2}{4}\Bigl)\\
R&=&\frac{1}{2}\Bigl[\Bigl(\frac{\omega_0^2}{4}-\frac{\Delta_0^2}{4}\Bigl)^2+4\hg^2\omega_0\Delta\Bigl]^\frac{1}{2}\no
\eea
The (transposed) eigenvector column $w_\lambda$ of $\theta^{-1}$ corresponding to eigenvalue $\lm$ $(\lm_\rho, \rho=1-4)$ is given  by
\be
w_\lm^T=\nu_{\lm}\Bigl(1,-\frac{2i\hg\frac{\Delta}{2}}{(\frac{\Delta}{2}+\lm)(\frac{\omega_0}{2}-\lm)},
-\frac{\frac{\Delta}{2}-\lm}{\frac{\Delta}{2}+\lm},-\frac{2i\hg\frac{\Delta}{2}}{(\frac{\Delta}{2}+\lm)(\frac{\omega_0}{2}+\lm)}
\Bigr)
\label{eq:eigen2}
\ee
with a normalisation constant 
\be
|\nu_\lm|^2=\frac{(\frac{\Delta}{2}+\lm)^2(\frac{\omega_0}{2}+\lm)^2(\frac{\omega_0}{2}-\lm)^2}
{4\frac{\Delta}{2}|\lm|[(\frac{\omega_0}{2}+\lm)^2(\frac{\omega_0}{2}-\lm)^2+4\hg^2\frac{\Delta}{2}\frac{\omega_0}{2}]}
\label{eq:eigen3}
\ee
The paraunitary transformation defines the vibronic normal mode coordinates $\gamma$ via $\eta=\theta^{-1}\gamma$ where 
$\eta=(\alpha^\da,\alpha)=(a^\da b^\da ab)$ are the original boson coordinates. Like the latter, the vibronic normal modes fulfill the bosonic commutation relations $[\gamma_\rho,\gamma_{\rho'}^\da]=I_{\rho\rho'}$. Explicitly we write $\gamma^\da=(\phi^\da,\psi^\da,\phi,\psi)$. They diagonalise the Hamiltonian in Eqs.~(\ref{eq:ham1},\ref{eq:ham2}) finally leading to
\be
H=\sum_n\Omega_\phi(\phi^\da_n\phi_n+\frac{1}{2})+
\sum_n\Omega_\psi(\psi^\da_n\psi_n+\frac{1}{2})
\ee
Where the triply degenerate (n=x,y,z) normal mode frequencies $\Omega_s^n =\Omega_s$ ($s=\phi,\psi$) are given by $\Omega_\phi=2\lambda_+$ and  $\Omega_\psi=2\lambda_-$. The relation between the non-interacting $\eta$-bosons and the new normal mode coordinates $\gamma$ is explicitly given by
\be
\left(\matrix{
a_n\cr
b_n\cr
a^\da_n\cr
b^\da_n}\right)=
\left(\matrix{
w^1_{\lm_1}&w^1_{\lm_2} & w^1_{\lm_3} &w^1_{\lm_4} \cr
w^2_{\lm_1}& w^2_{\lm_2}& w^2_{\lm_3}& w^2_{\lm_4}\cr
w^3_{\lm_1}& w^3_{\lm_2}& w^3_{\lm_3}& w^3_{\lm_4} \cr
w^4_{\lm_1}& w^4_{\lm_2} & w^4_{\lm_3}&w^4_{\lm_4}}\right)
\left(\matrix{
\phi_n\cr
\psi_n\cr
\phi^\da_n\cr
\psi^\da_n}\right)
\label{eq:paratrans}
\ee
where the matrix elements $w^k_{\lambda_j}$ are obtained from Eqs.~(\ref{eq:eigen2},\ref{eq:eigen3}) using the eigenvalues in Eq.~(\ref{eq:eigen1}).

%
\begin{figure}
\includegraphics[width=8cm]{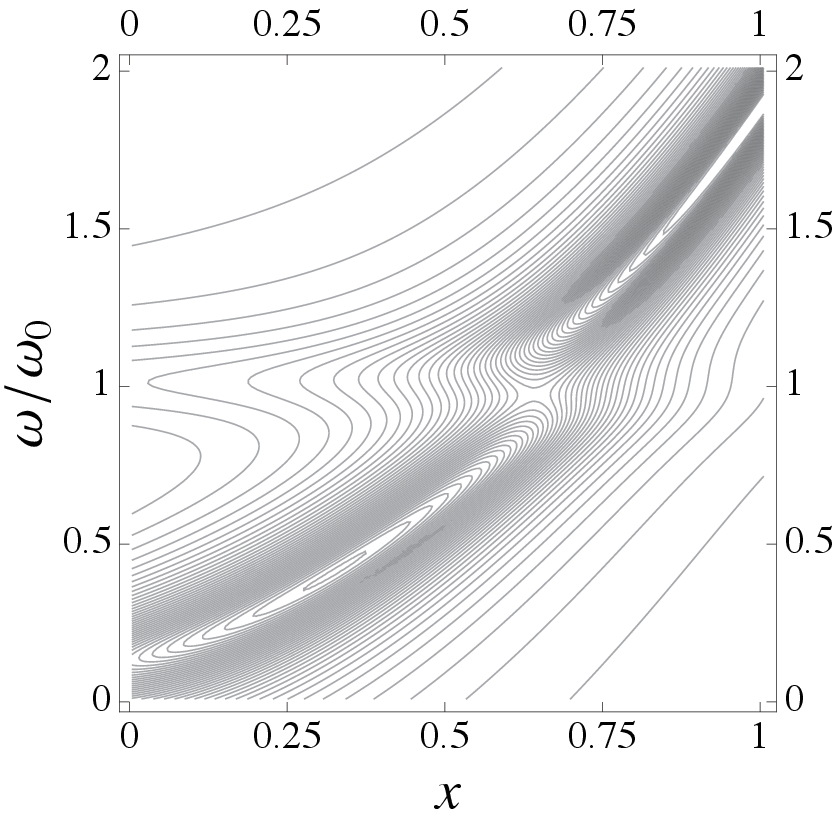}\hfill
\includegraphics[width=8cm]{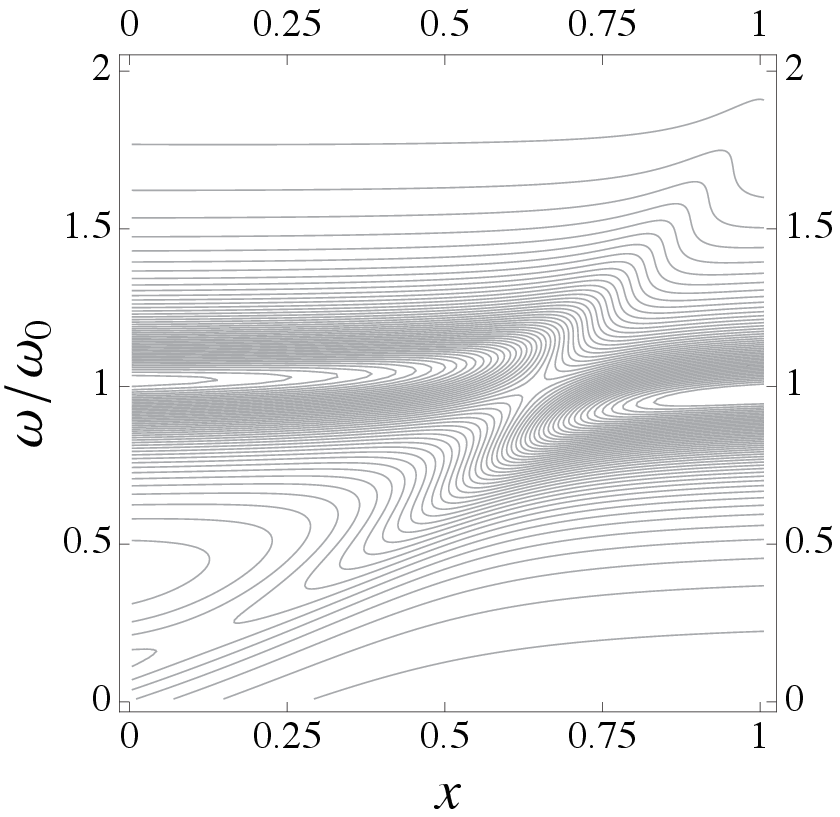}
\caption{Contour plots of dipolar $S_D(\omega)$ (left) and rattling phonon $S_r(\omega)$ (right) spectral functions. 
For better visibility of the vibronic effect we choose the coupling constant $\gamma =0.3$ and a
line width $\Gamma_s/\omega_0 =0.15$.}
\label{fig:Fig2}
\end{figure}
%

\section{Dynamical susceptibilities and structure functions}
\label{sec:dynsus}

With the above closed solution all interesting susceptibilities and dynamical structure functions of the model may be calculated analytically. The dipolar structure function is proportional to the INS cross section and is therefore the most direct means to observe the vibronic modes. Furthermore they may influence the NMR rate which is obtained from the dynamical dipolar susceptibility.
The dynamics of quadrupole moments and the phonon spectral functions may be obtained in a similar way.

\subsection{Dipolar susceptibility and INS spectral function}
\label{subsec:dipolar}

Using the representation in Eq.~(\ref{eq:dipolar}) for the dipole operator the dipolar susceptibility may be expressed as
\be
\chi_D(\omega)=ib_D^2\int_{-\infty}^{\infty}
dte^{i\omega t}\la[(a+a^\da)_t,(a+a^\da)_0]\ra\theta_H(t)
\ee
where $\theta_H(t)$ is the Heaviside function. Applying the paraunitary transformation we obtain
\bea
a+a^\da&=&u_{D\phi}(\phi+\phi^\da)+u_{D\psi}(\psi+\psi^\da)
\eea
with $u_{Ds}$ ($s=\phi,\psi$) given by
\bea
u_{Ds}^2=\frac{\Omega_s}{\Delta}
\frac{(\omega_0^2-\Omega_s^2)^2}{(\omega_0^2-\Omega_s^2)^2+\gamma^2\Delta\omega_0}
\label{eq:uD}
\eea
where we defined $\gamma =4\hat{g}_0 = 4b_Qg_0(2M\omega_0)^{-\frac{1}{2}}$. This leads to the dynamical dipolar susceptibility 
\be
\chi_D(\omega)=-b_D^2\sum_s u_{Ds}^2D_s(\omega)
\label{eq:chiD}
\ee
where $D_s(\omega)$ is the retarded Greens function of normal mode bosons according to
\be
D_\phi(\omega)=i\int_{-\infty}^{\infty}
dte^{i\omega t}\la[(\phi+\phi^\da)_t,(\phi+\phi^\da)_0]\ra\theta_H(t)=
\frac{2\Omega_\phi}{(\omega+i\delta)^2-\Omega_\phi^2}
\ee
and a similar equation for $D_\psi(\omega)$. The corresponding spectral function is given by
\be
\hat{S}_s(\omega)=-\frac{2}{1-e^{-\beta\omega}}D_s^{''}(\omega)
\ee
Using the fluctuation dissipation theorem $\hat{S}_s(-\omega) = e^{-\beta\omega}  \hat{S}_s(\omega)$  ($\beta=1/kT$) the total
dipolar spectral function corresponding to Eq.~(\ref{eq:chiD}) is then obtained as
\be
S_D(\omega)=2\pi b_D^2\sum_s
\frac{\Omega_s}{\Delta}
\frac{(\omega_0^2-\Omega_s^2)^2}{(\omega_0^2-\Omega_s^2)^2+\gamma^2\Delta\omega_0}
\Bigl[(n_s+1)\delta(\omega-\Omega_s)+n_s\delta(\omega+\Omega_s)\Bigr]
\label{eq:strucD}
\ee
Here $n_s=(e^{\beta\Omega_s}-1)^{-1}$ is the Bose distribution function. 
When we include a constant finite linewidth $\Gamma_s$ for the $\phi, \psi$ bosons the delta functions have to be
replaced by Lorentzians. Then at zero temperature we obtain
\be
S_D(\omega)=2\pi b_D^2\sum_s
\frac{\Omega_s}{\Delta}
\frac{(\omega_0^2-\Omega_s^2)^2}{(\omega_0^2-\Omega_s^2)^2+\gamma^2\Delta\omega_0}\cdot
\frac{\Gamma_s/\pi}{(\omega-\Omega_s)^2+\Gamma_s^2}
\label{eq:strucD0}
\ee
Finally, using $D'_s(0)=-(2/\Omega_s)$ we obtain the zero-temperature static dipolar susceptibility as
\be
\chi_D(\omega =0)=\Bigl(\frac{2b_D^2}{\Delta}\Bigr)\sum_s
\frac{(\omega_0^2-\Omega_s^2)^2}{(\omega_0^2-\Omega_s^2)^2+\gamma^2\Delta\omega_0}
\equiv \frac{2b_D^2}{\Delta}
\label{eq:suscD}
\ee
Therefore the static dipolar susceptibility of the vibronic system for $\gamma \neq 0$ is unchanged from the dipolar van-Vleck susceptibility of the uncoupled ($\gamma =0$) singlet-triplet CEF states because the formation of vibronic modes  involves only the quadrupolar CEF excitations. The result in Eq.~(\ref{eq:strucD0}) together with Eq.~(\ref{eq:eigen1}) gives the frequency and $x, \gamma$ dependence of the dipolar spectral function which may be compared with INS results. This will be further discussed in Sec.~\ref{subsec:specnum}.

%
\begin{figure}
\includegraphics[width=8cm]{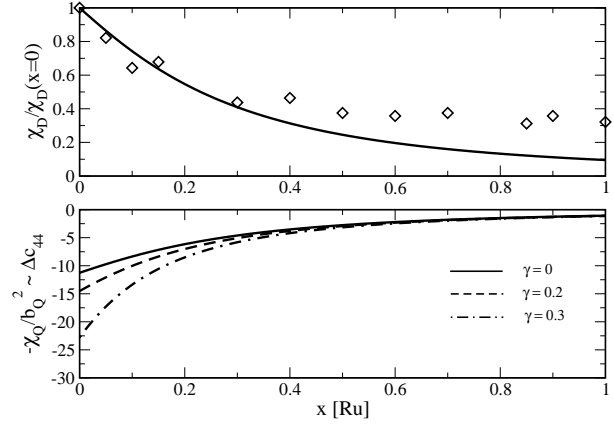}
\caption{The top panel shows the calculated T=0 dipolar susceptibility (which is independent of vibronic coupling $\gamma$) in comparison to experimental values from Ref.~\onlinecite{Frederick04}.  Bottom panel shows the ($\gamma$ -dependent) T=0 quadrupolar susceptibility. It is proportional to the T=0 elastic constant reduction $\Delta c_{44}$.}
\label{fig:Fig3}
\end{figure}
%

\subsection{Quadrupolar susceptibility and rattling phonon spectral function}
\label{subsec:quadrupolar}

The dynamical quadrupolar susceptibility, the phonon Green's function and their associated spectral function may be obtained in a completely analogous way. Using the bosonic representation of quadrupolar operators in Eq.~(\ref{eq:quadru}) we have
\be
\chi_Q(\omega)=-ib_Q^2\int_{-\infty}^{\infty}
dte^{i\omega t}\la[(a-a^\da)_t,(a-a^\da)_0]\ra\theta_H(t)
\ee
Again replacing the a- bosons with $\phi,\psi$  bosons by the paraunitary transformation we get
\bea
a-a^\da&=&u_{Q\phi}(\phi-\phi^\da)+u_{Q\psi}(\psi-\psi^\da)
\eea
where now we have a slightly different
\bea
u_{Qs}^2=\frac{\Delta}{\Omega_s}
\frac{(\omega_0^2-\Omega_s^2)^2}{(\omega_0^2-\Omega_s^2)^2+\gamma^2\Delta\omega_0}
\label{eq:uQ}
\eea
Similar as before the quadrupolar spectral function is obtained as
\be
S_Q(\omega)=2\pi b_Q^2\sum_s
\frac{\Delta}{\Omega_s}
\frac{(\omega_0^2-\Omega_s^2)^2}{(\omega_0^2-\Omega_s^2)^2+\gamma^2\Delta\omega_0}
\Bigl[(n_s+1)\delta(\omega-\Omega_s)+n_s\delta(\omega+\Omega_s)\Bigr]
\label{eq:strucQ}
\ee
and for finite boson line width and in the limit T=0 we likewise obtain 
\be
S_Q(\omega)=2\pi b_Q^2\sum_s
\frac{\Delta}{\Omega_s}
\frac{(\omega_0^2-\Omega_s^2)^2}{(\omega_0^2-\Omega_s^2)^2+\gamma^2\Delta\omega_0}\cdot
\frac{\Gamma_s/\pi}{(\omega-\Omega_s)^2+\Gamma_s^2}
\label{eq:strucQ0}
\ee
Furthermore the static zero-temperature quadrupolar  susceptibility may be obtained as
\be
\chi_Q(0)=\Bigl(\frac{2b_Q^2}{\Delta}\Bigr)\sum_s\Bigl(\frac{\Delta}{\Omega_s}\Bigr)^2
\frac{(\omega_0^2-\Omega_s^2)^2}{(\omega_0^2-\Omega_s^2)^2+\gamma^2\Delta\omega_0}
\label{eq:suscQ}
\ee
The prefactor is the quadrupolar van-Vleck susceptibility of the uncoupled singlet-triplet states.
The quantity in Eq.~(\ref{eq:suscQ}) depends on $\gamma$ and therefore on the mode splitting, contrary to $\chi_D(0)$. It  is in principle accessible in ultrasonic experiments where it determines the velocity or 
elastic constant change for T $\rightarrow$ 0. The latter is given by $\Delta c_{44}(T \rightarrow 0) = -g_{44}^2\chi_Q$  where
$g_{44}$ is the magnetoelastic coupling constant of the $c_{44}$ transverse mode  propagating along (001) direction \cite{Thalmeierbook}.\\

Now we consider the retarded propagator of the rattling phonon which is defined by
\be
D_r(\omega)=-i\int_{-\infty}^{\infty}
dte^{i\omega t}\la[(b+b^\da)_t,(b+b^\da)_0]\ra\theta_H(t)
\ee
Applying the paraunitary transformation we may express
\bea b+b^\da&=&u_\phi(\phi-\phi^\da)+u_\psi(\psi-\psi^\da)
\eea
where $u_s=i|u_s|$ and the modulus is now given by
\bea
|u_s|^2=\frac{\omega_0}{\Omega_s}
\frac{\gamma^2\Delta\omega_0}{(\omega_0^2-\Omega_s^2)^2+\gamma^2\Delta\omega_0}
\eea
This leads to a spectral function of the rattling phonon propagator
\be
S_r(\omega)=2\pi\sum_s
\frac{\omega_0}{\Omega_s}
\frac{\gamma^2\Delta\omega_0}{(\omega_0^2-\Omega_s^2)^2+\gamma^2\Delta\omega_0}
\Bigl[(n_s+1)\delta(\omega-\Omega_s)+n_s\delta(\omega+\Omega_s)\Bigr]
\label{eq:strucr}
\ee
Including the finite line width $\Gamma_s$  for the normal modes we obtain the zero temperature limit
\be
S_r(\omega)=2\pi\sum_s
\frac{\Delta}{\Omega_s}
\frac{\gamma^2\Delta\omega_0}{(\omega_0^2-\Omega_s^2)^2+\gamma^2\Delta\omega_0}\cdot
\frac{\Gamma_s/\pi}{(\omega-\Omega_s)^2+\Gamma_s^2}
\label{eq:strucr0}
\ee

The rattling phonon spectral function in Eq.~(\ref{eq:strucr0}) is complementary to the dipolar and quadrupolar spectral function.
In our localized model they are momentum independent.
However in the INS cross section the latter is multiplied by the square of the electronic form factor $F(\vQ)$ of the 4f shell which decreases with $|\vQ|$  while the former is multiplied by $|\vQ|^2$. Therefore the dipolar excitation may be seen at small and the rattling phonon part at large total momentum transfer.

\subsection{Numerical results for spectral function and static susceptibilities}
\label{subsec:specnum}

The basic feature of the vibronic mode formation is shown in Fig.~\ref{fig:Fig1}. At the crossing of the bare ($\gamma$=0, dashed lines) rattling mode $\omega_0(x)$ and $\Delta_0(x)$ a repulsion takes place for finite magnetoelastic coupling $\gamma$  and anti-crossing mixed modes (full lines) $\Omega_\phi$ (upper mode) and $\Omega_\psi$ (lower mode) are formed. Their splitting increases with coupling strength. At the crossing where $\omega_0(x_c)=\Delta(x_c)$ we have $\delta = \Omega_\phi -\Omega_\psi \simeq \gamma$ as long as $\gamma/\omega_0 \ll 1$. For $\gamma = 0.25$ the splitting is still moderate enough to be compatible with the mode energies determined from specific heat analysis \cite{Miyazaki09}. The determination of the mode splitting and $\Omega_s(x)$ ($s=\phi,\psi$) requires the investigation of the dynamical magnetic and phononic structure function in INS experiments. The former should be proportional to $S_D(\omega)$ and the latter to $S_r(\omega)$. These functions are calculated from Eqs.~(\ref{eq:strucD0},\ref{eq:strucr0}), respectively, and are shown in Fig.~\ref{fig:Fig2}. Away from the anti-crossing region $S_D(\omega)$ has appreciable intensity only around the bare CEF excitation $\Delta(x)$  and  $S_r(\omega)$ only around the bare rattling phonon frequency $\omega_0$. In the anti-crossing region  $S_{D,r}(\omega)$ have equal intensity at both split modes $\Omega_s(x)$. Observation of this feature by future INS experiments would directly confirm the vibronic mixed mode formation in \POx.\\

The static dipolar and quadrupolar susceptibilities $\chi_{D,r}(\omega)$ are also accessible in experiments. However the former does not show an effect of the mode coupling $\gamma$ but remains the unrenormalised singlet-triplet van Vleck susceptibility (Eq.~(\ref{eq:suscD})). Therefore no information on the mode splitting can be gained from it. The behaviour of $\chi_D(x)$  (T $\rightarrow$ 0), using a constant matrix element $b_D$ in comparison to experimental data from Ref.~\onlinecite{Frederick04} is shown in the top Fig.~\ref{fig:Fig3}. Up to $x\simeq 0.4$ the behaviour is in agreement however for larger x the experimental values show no further decrease. This may be in part due to the increasing relative importance of higher levels when the singlet-triplet $\Delta(x)$ increases (Fig.~\ref{fig:Fig1}). Furthermore if the tetrahedral CEF part increases with x the dipolar matrix element $b_D \sim d$ will also increase leading to a larger $\chi_D$ at higher x. There is evidence from the superconducting pair breaking behaviour discussed in the next section that this is indeed the case.
The quadrupolar ($\Gamma_5$-type) susceptibility may be obained from the T=0 suppression of the appropriate (c$_{44}$) symmetry elastic constant. The suppression gets larger for increasing vibronic coupling $\gamma$ at small x. Therefore measuring $\Delta c_{44}(T\rightarrow 0;x)$ may be used as an indirect means to determine the coupling strength.

\section{Superconducting pair forming  and pair breaking by vibronic excitations}
\label{sec:crittemp}

The effective pairing interaction for the formation of Cooper pairs in skutterudites consists of three contributions: (i) harmonic phonons, (ii) local CEF excitations. (iii) low energy rattling (anharmonic) phonons. The former two have been included in the model for \POy~ \cite{Chang07} (y = Pr-concentration) and the latter were proposed for the pyrochlore superconductor KOs$_2$O$_6$ in Ref. \onlinecite{Chang09}.  However the NMR relaxation \cite{Nakai08}, ultrasonic experiments \cite{Goto05} and specific heat measurements \cite{Miyazaki09} have indicated that rattling phonons are also present in La, Pr-skutterudite compounds and therefore may contribute to the effective pairing mechanism. In fact the INS experiments in CeRu$_4$Sb$_{12}$ have shown \cite{Lee06} the existence of a flat optical phonon mode at $\omega_0$ = 6 meV within the acoustic phonon band which shows little hybridisation with the latter. The flat optical mode was interpreted as a (Ce-) guest (rattling) mode within the rigid cage structure formed by Sb$_{12}$ icosahedrons. This mode  is anharmonic and the effective frequency $\omega_e(T)$ is temperature dependent. Around T$_c$ however this may be neglected since $\omega_e(T)$ has reached its low temperature asymptotic value  $\omega_0$. In the case of \POy~(x=0) which was studied in Ref.~\onlinecite{Chang07} one has $\Delta\ll\omega_0$, therefore the interaction of rattling phonon and singlet-triplet excitation may be neglected. This may not hold in \POx~ for general x because $\Delta(x)$ crosses $\omega_0$ at $x_c\sim 0.65$. Therefore the nonadiabatic vibronic spectral function should be used in modeling the effective pairing interaction for arbitrary Ru content x. We mention that 'nonadiabatic' refers to the localised 4f electron- phonon interaction, not to the conduction electron- phonon term.\\ 

%
\begin{figure}
\includegraphics[width=8cm]{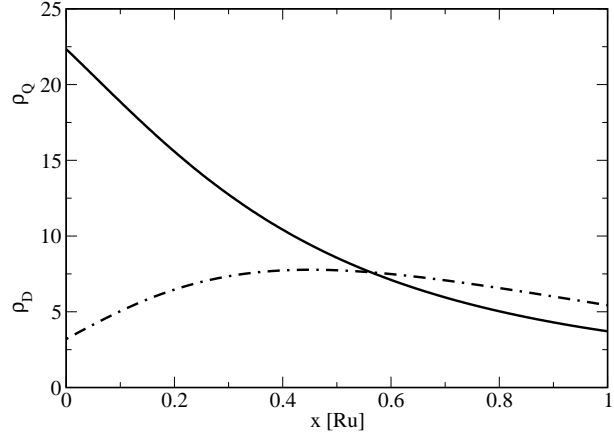}
\caption{Crossover from pair forming (T$_c >$ T$_c^0$) at x =0 with $\rho_Q > \rho_D$ to pair breaking at x=1
with $\rho_Q < \rho_D$  (T$_c <$ T$_c^0$). At x=0 $\rho_D/\rho_Q =1/7$ as in Ref.~\onlinecite{Chang07} which produces the experimental enhancement $T_c/T_c^0$=2.5. At x=1 $\rho_D/\rho_Q = 1.43$ is adjusted to the experimental  depression $T_c/T_c^0$=0.34.}
\label{fig:Fig4}
\end{figure}
%

Experimentally it was found early in Ref.~\onlinecite{Frederick04} that T$_c$(x) has a minimum close to x$_c$ $\simeq$ 0.65. Therefore the question arises whether this is tied to a suggested vibronic mode splitting or to a different origin. The most convenient starting point for a theoretical description is the T$_c^0(x)$ background variation in \LOx~ which is determined by the harmonic and rattling phonon mechanism. The microscopic model behind will not be further specified. The symmetry of the superconducting order parameter in the skutterudites  is presumably of (anisotropic) extended s-wave type \cite{Parker08,Chang07}. For the present purpose we ignore the superconducting gap anisotropy. The presence of 4f states in \POx~ leads to a scattering of conduction electrons from singlet-triplet CEF excitations. This modifies the pair amplitude and changes the background T$_c^0(x)$ of the La compound to a renormalised  T$_c(x)$. This process is due to exchange and aspherical Coulomb scattering of conduction electrons from the 4f shell given by \cite{Chang07}
\be
H_{sf}=-I_{ac}\sum_{\vq\vk,n\sigma}f_n(\vq)O_{\vq n}c^\dagger_{\vk\sigma}c_{\vk+\vq\sigma} 
-(g_J-1)I_{ex}\sum_{\vq\vk,n\sigma}\sigma^{\sigma\sigma'}_nJ_{\vq n}c^\dagger_{\vk\sigma}c_{\vk+\vq\sigma'}
\label{eq:hsf}
\ee
Here $O_{\vq n}=\sum_\vq O_n(i)\exp(i\vq\vR_i)$ (n=yz,zx,xy) and  $J_{\vq n}=\sum_\vq J_n(i)\exp(i\vq\vR_i)$ (n=x,y,z) are quadrupolar and dipolar operators, respectively. Furthermore $g_J=4/5$ is the Land\'e factor and $f_{n\vq} =\hq_y\hq_z, \hq_z\hq_x, \hq_x\hq_y$ are quadrupolar form factors ($\hat{\vq}=\vq/|\vq|$).
The principal effect of $H_{sf}$ on superconducting properties of 4f compounds with CEF splitting  has been investigated in Ref.~\onlinecite{Fulde70}. It was found that for singlet superconductors aspherical Coulomb (quadrupolar) scattering which supports pair formation and enhances T$_c^0$ because $O_n$ (Eq.~(\ref{eq:quadru})) is even under time reversal. In contrast the exchange term leads to pair breaking and reduces the background  T$_c^0$ because $J_n$  is odd under time reversal. For the case of having only a twofold Kramers degenerate ground state level the latter is described by the well known Abrikosov-Gorkov \cite{Abrikosov60} theory. The modified T$_c$(x) of \POx~ includes both effects because the singlet-triplet excitations have dipolar as well as quadrupolar character due to the tetrahedral CEF (Sec.~\ref{sec:model}). In the present case their magnetoelastic interaction with rattling phonons leads to vibronic excitation modes with modified dipolar and quadrupolar matrix elements. Generalization of the expressions for pure CEF systems in Refs.~(\onlinecite{Fulde70,Chang07}) to vibronic excitations leads to an equation for the renormalised $T_c$ given by
\be
-\frac{8}{\pi}\Bigl(\frac{T_c}{T^0_c}\Bigr)\ln\Bigr(\frac{T_c}{T^0_c}\Bigr)=
\rho_Q\sum_su_{Qs}^2(\Omega_s)F(\frac{\Omega_s}{2T_c})+\rho_D\sum_su_{Ds}^2(\Omega_s)G(\frac{\Omega_s}{2T_c})
\label{eq:Tc}
\ee
Here $u_{Ds}^2, u_{Qs}^2$ are given by Eqs.~(\ref{eq:uD},\ref{eq:uQ}). The dimensionless vibronic pair forming and breaking strengths $ \rho_Q(x)$ and $\rho_D(x)$ for  the quadupolar and dipolar conduction electron scattering channels due to $H_{sf}$ are given by 
\bea
\rho_Q(x)&=&y\hat{\rho}_Q(x)\frac{b_Q^2(x)}{b_Q^2(0)}\frac{1}{t_c^0(x)} ;\qquad 
\hat{\rho}_Q(x)=\frac{2\pi N_FI_{ac}^2\la f^2\ra}{T_c^0(0)}3b_Q^2(0)\no\\
\rho_D(x)&=&y\hat{\rho}_D(x)\frac{b_D^2(x)}{b_D^2(0)}\frac{1}{t_c^0(x)} ;\qquad 
\hat{\rho}_D(x)=\frac{2\pi N_FI_{ex}^2(g_J-1)^2}{T_c^0(0)}3b_D^2(0)
\label{eq:scatt}
\eea
where y is the Pr content. In the present case of \POx~  we have $y=1$.  Furthermore N$_F$ is the conduction electron DOS of \LOx~ at the Fermi level and $t_c^0(x)=T_c^0(x)/T_c^0(0)$ the normalised background transition temperature of \LOx~  ($y=0$) with $T_c(0)=0.74$ K.
The quadrupolar and dipolar matrix elements $b_Q, b_D$ are given in Appendix \ref{sec:A1} and $\la f^2\ra$ is the  Fermi surface averages of the quadrupolar form factors (independent of n). The functions F, G in Eq.~(\ref{eq:Tc}) which correspond to pair formation and pair breaking respectively in principle depend explicitly on temperature via the thermal occupation of excited vibronic levels \cite{Fulde70}. However in the present case we have $\Delta(x)/2T^0_c(x)\gg 1$ and therefore $\Omega_s(x)/2T_c^0(x) \gg 1$ for all Ru concentrations x. Then we may use the low temperature limit for F, G in which case we have, defining $x_s=\Omega_s/2T_c$:
\bea
F(x_s)&=&-\frac{1}{x_s}+S_1(x_s)\no\\
G(x_s)&=&\frac{1}{x_s}-S_1(x_s)+S_2(x_s)=-F(x_s)+S_2(x_s)
\label{eq:FGfunc}
\eea
Here $S_{1,2}(x_s)$ are combinations of digamma functions derived in Ref.~\onlinecite{Fulde70} and given in Appendix \ref{sec:A2} for completeness. 
The dimensionless $\rho_Q(x),\rho_D(x)$ parameters in Eqs.~(\ref{eq:Tc},\ref{eq:scatt})  depend on the Ru content x via three factors: (i) the quadrupolar and dipolar matrix elements $b_i(x)$ ($i=Q,D$) which are determined by the x-dependent CEF parameters $x_{CF}(x)$ and $y_{CF}(x)$ (Appendix \ref{sec:A1}). (ii) the dimensionless interaction constants $\hat{\rho}_i(x)$  depend on x through conduction electron DOS N$_F$ and possibly also through the interaction strengths $I_{ac}, I_{ex}$. (iii) the background normalised transition temperature $t^0_c(x)$.\\

%
\begin{figure}
\includegraphics[width=8cm]{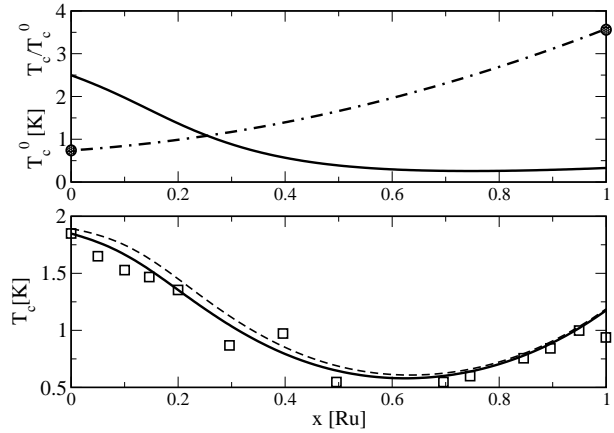}
\caption{Dependence of critical temperature on x (Ru-content). {\it Top}: Dash-dotted line shows interpolated T$_c^0(x)$ curve for \LOx. Only T$_c^0(0)$ = 0.74 K and  T$_c^0(1)$ = 3.58 K (full circles) are experimentally known  values. Full line shows the calculated renormalisation factor T$_c(x)$/T$_c^0(x)$ (with $\gamma = 0$). Note that  T$_c^0(0) < $ T$_c(0) = 1.85$ K but  T$_c^0(1) >$ T$_c(1) = 1.20$ K. {\it Bottom}: T$_c(x)$ for \POx. Full line is obtained from top panel. The experimental T$_c$ data (squares) are taken from susceptibility results in Ref.~\onlinecite{Frederick04}. The minimum is due to a decreasing T$_c(x)$/T$_c^0(x)$ and an increasing  T$_c^0(x)$. Dashed line ($\gamma =0.15$) shows that effect of vibronic level splitting on T$_c$(x) is small.}
\label{fig:Fig5}
\end{figure}
%
A model for these Ru concentration dependences is needed as input to calculate the T$_c$(x) for the Pr compound from  Eq.~(\ref{eq:Tc}) or T$^0_c(x)$/T$^0_c(0)$ normalised to the \LOx~ value. The model for $b_i(x)$ ($i=Q,D$) and $\hat{\rho}_i(x)$  is described in Appendix \ref{sec:A1}. The background T$^0_c(x)$ is determined experimentally only for the stoichiometric cases ($x=0,1$). The interpolation in Fig.~\ref{fig:Fig5} is used for intermediate concentrations. 
The resulting x-dependence of pair-forming and breaking parameters $\rho_Q(x)$ and $\rho_D(x)$ obtained from Eq.~(\ref{eq:scatt}) is shown in Fig.~\ref{fig:Fig4} and fit procedure and parameters are explained in Appendix \ref{sec:A1} and in  the caption. 
The decrease in $\rho_Q$ is mostly due to the increase of $t_c^0(x)$ =  T$^0_c(x)$/T$^0_c(0)$ (dash-dotted curve in Fig.~\ref{fig:Fig5}). On the other hand $\rho_D$ has to increase with x in order to achieve the crossover from T$_c$ enhancement (T$_c$(0)/T$^0_c$(0)=2.5) at x=0 to T$_c$ depression (T$_c$(1)/T$^0_c$(1)=0.34) at x=1. This observed crossover from  pair forming to pair breaking by CEF excitations is only possible if the dipolar scattering strength increases with x. This is in part due to the increase of the tetrahedral CEF parameter $y_{CF}$ (Appendix~\ref{sec:A1}) since for small  $y_{CF}$ the dipolar matrix element $d\sim y_{CF}^2$ shows a strong increase with $y_{CF}$. The latter is compensated by the $t_c^0(x)$ increase leading to an almost flat $\rho_D(x)$ for large x. The associated crossover from T$_c(x)$/T$_c^0$ enhancement to reduction obtained from Eq.~(\ref{eq:Tc}) is shown in Fig.~\ref{fig:Fig5}  with an almost flat reduction factor at large x. Altogether, because the background $T_c^0(x)$ strongly increases with x, the renormalised T$_c(x)$ exhibits a minimum for intermediate concentrations (bottom Fig.~\ref{fig:Fig5}). The experimental values from Ref.~\onlinecite{Frederick04} are shown for comparison (squares). The T$_c(x)$ curve was calculated for $\gamma =0$ and moderate $\gamma=0.15$ with only small difference, especially for larger x (around the minimum region). The precise form of  $\rho_Q(x)$ and $\rho_D(x)$ cannot be determined presently because no reliable information on CEF parameters and matrix elements for intermediate x is available. However the crossover from mainly pair forming at small x to pair breaking behaviour at large x (Fig.\ref{fig:Fig4}) is robust.
We conclude that the vibronic splitting does not play an essential role in the T$_c$(x) minimum formation. This is due to the fact that for $x\simeq x_c$ close to the crossing region already $\Omega_s(x_c)/2T_c \gg 1$ where the pair breaking functions $S_{1,2}(x_s)$ (Appendix~\ref{sec:A2}) vary slowly with $x_s$. Therefore the coupling $\gamma > 0$ hardly affects T$_c$ for larger x.  Its effect would be much bigger if the mode crossing  would appear at energies comparable to T$_c$, i.e.   $\Omega_s(x)/2T_c \simeq 1$. In the case of \POx~ this is not possible because already for x=0 we have $\Delta_0/2T_c^0 =5.4 $. \\

One may conclude that the $T_c$(x) minimum is not directly linked to the crossing of rattling phonon mode and CEF excitation found in Ref.~\onlinecite{Miyazaki09}. It is rather a combined effect involving the crossover from T$_c$/T$_c^0$ enhancement to reduction and and the increase in the background T$_c^0(x)$. In this scenario the observed  $T_c$(x) minimum also does not imply or suggest a symmetry change of the order parameter from below to above the Ru concentration at the minimum.

\section{Summary and Conclusion}
\label{sec:conc}

In this work the possible nonadiabatic effects of CEF singlet-triplet excitations and rattling phonons of rare earth hosts in the cages of \POx~  have been investigated. This is suggested by specific heat experiments of Miyazaki et al \cite{Miyazaki09} which show a crossing of triplet excitation and rattling phonon energies at an intermediate Ru content.\\

It has been proposed that a magnetoelastic coupling between the singlet-triplet excitations and the local rattling modes should lead to vibronic splitting and mixed-mode formation around the crossing point. These features can be detected in the spectral functions measured by INS experiments. It should also be observable in the low temperature depression of the symmetry elastic constant as function of Ru concentration which measures directly the quadrupolar susceptibility of the vibronic excitations. On the other hand the magnetic susceptibility is not affected by the mode splitting. Its comparison with experiment indicates an increase of dipolar matrix elements for increasing x and possibly the influence of the higher lying triplet $\Gamma_4^{(1)}$.\\ 

The superconducting T$_c$(x) behaviour of \POx~ shows an enhancement at small  and reduction at larger Ru concentration compared to the background T$^0_c$(x) of \LOx. It also exhibits a minimum at intermediate concentration. The analysis presented here suggest that this behaviour is the result of a crossover between primarily quadrupolar pair formation and mainly dipolar pair breaking mechanism originating in the singlet-triplet excitations. The vibronic coupling has little influence on the existence of the minimum (Fig.~\ref{fig:Fig5}) because of the large singlet-triplet splitting for $x \simeq 0.65$ in comparison to T$_c$ . The minimum is rather a combined effect of a decreasing enhancement factor and the increasing background T$^0_c$(x). An essential feature of the model is a growing dipolar (magnetic) character of the triplet for larger Ru content which may be due to an increase of the tetrahedral crystal field part. For a better determination of the model parameters it would therefore be important to determine the CEF parameters for intermediate Ru content by INS experiments. The experimental knowledge of $T_c^0(x)$ of \LOx~ in the whole Ru concentration range would also allow further improvement of the model.


\appendix

\section{}
\label{sec:A1}
In this appendix we summarize basic properties of the tetrahedral CEF model and interaction model needed in the analysis.
The CEF potential in tetrahedral symmetry is given in Ref. \onlinecite{Takegahara01}. Aside from an overall scale W it is determined by two parameters x$_{CF}(x)$, y$_{CF}(x)$ which characterize fourfold cubic and tetrahedral contributions respectively. These parameters will depend on the Ru content x. The same is then true for dipolar and quadrupolar matrix elements $b_D^2(x)=(20/3)d^2(x)$ and $b_Q^2(x)=(140/3)(1-d^2(x))$ which depend on a mixing parameter d given by \cite{Shiina04a}
\bea
d^2(x)=\frac{1}{2}\Bigl(1-\frac{3+2x_{CF}(x)}{(3+2x_{CF}(x))^2+1008y_{CF}(x)^2}\Bigr)
\label{eq:CEFmix}
\eea
It is a measure of the tetrahedral CEF since $d=0$ for $y_{CF}=0$. In this model the singlet-triplet $\Gamma_1-\Gamma_4^{(2)}$ splitting is given by \cite{Shiina04a}
\bea
\Delta(x)=2W(36-58x_{CF}(x))-4W\bigl[(3+2x_{CF}(x))^2+1008y^2_{CF}(x)\bigr]^\frac{1}{2}
\label{eq:CEFsplit}
\eea
For \PO~ (x=0) we use x$_{CF}(0)$=0.45 and $y_{CF}(0)= 0.1$  \cite{Shiina04a}. In \PR~ (x=1) the CEF splitting is much larger. This suggests that $x_{CF}(1)$ is close to zero according to the LLW tables \cite{Lea62}. Furthermore for $x = 1$ the dipolar matrix element $b_D^2$ and hence $y_{CF}$ has to increase. Therefore we use the set  x$_{CF}(1)$ = 0.0 and $y_{CF}(1)= 0.2$. It leads to comparable intensities for the $\Gamma_1\rightarrow\Gamma_4^{(2)} $ (84 K) and 
 $\Gamma_1\rightarrow\Gamma_4^{(2)} $ (145 K) transitions in qualitative agreement with INS results in Ref.~\cite{Adroja05}. Furthermore the  CEF scale factor in Eq.(~\ref{eq:CEFsplit}) is given by W $\simeq$ 1.9 K leading to $\Delta(0)=8$ K and $\Delta(1)=84$ K. For general x an interpolation between values at $x=0,1$  is employed according to
\bea
x_{CF}(x)=x_{CF}^0(1-x)+x_{CF}^1x+x_{CF}^2x(1-x)
\label{eq:CEFinter}
\eea
and similar for $y_{CF}(x)$. The interpolation parameter sets $(x_{CF}^0,x_{CF}^1,x_{CF}^2)=(0.45,0,0)$ and $(y_{CF}^0,y_{CF}^1,y_{CF}^2)=(0.1,0.2,0.1)$ have been used for calculating $d^2(x)$ from Eq.~(\ref{eq:CEFmix}) and then $b_i^2(x))$ ($i=Q,D$) needed in Eq.~(\ref{eq:scatt}) . These CEF parameters also reproduce the experimental $\Delta(x)$ behaviour.

For the calculation of pair forming and breaking functions in Eq.~(\ref{eq:scatt}) one needs  the dimensionless interaction constants $\hat{\rho}_i(x)$ $(i=Q,D)$ in additon to the CEF matrix elements. The former are obtained by a similar interpolation as in  Eq.~(\ref{eq:CEFinter}) with parameter sets $(\hat{\rho}_Q^0,\hat{\rho}_Q^1,\hat{\rho}_Q^2)=(22.34,22.34,0)$ and $(\hat{\rho}_Q^0,\hat{\rho}_Q^1,\hat{\rho}_Q^2)=(3.2,10.3,6.0)$. The $\rho_i^{0,1)}$ are chosen such that for $x=0,1$ the absolute value of $T_c(x)$  corresponds to the experimental one in Fig.~\ref{fig:Fig5}. Finally the $\hat{\rho}_i^2$ are adjusted to obtain the minimum position and value of  $T_c(x)$. 

\section{}
\label{sec:A2}

The functions $S_{1,2}(x)$ in Eq.~(\ref{eq:FGfunc}) were derived in Ref.~\onlinecite{Fulde70} and are given here for completeness:
\bea
S_1(x)&=&\frac{4x}{\pi^4} Re\sum_{n=0}^\infty\frac{1}{(n+\frac{1}{2})}\frac{1}{(n+\frac{1}{2}-\frac{ix}{\pi})^2}
\bigl[\psi(1+n-\frac{ix}{\pi})-\psi(\frac{1}{2})\bigr] \no\\
S_2(x)&=&\frac{8}{\pi^3} Im\sum_{n=0}^\infty\frac{1}{(n+\frac{1}{2}-\frac{ix}{\pi})^2}
\bigl[\psi(1+n-\frac{ix}{\pi})-\psi(\frac{1}{2})\bigr] 
\eea
where $\psi(z)=d\ln\Gamma(z)/dz$ is the digamma function \cite{Abramowitzbook}

\section*{Acknowledgements}
The author would like to thank Y. Aoki and H. Sato for discussion and for drawing his attention to the subject.

\end{document}